\documentclass[11pt]{article}

\usepackage[nonatbib, final]{neurips_2021}

\usepackage{float}
\usepackage[utf8]{inputenc} 
\usepackage[T1]{fontenc}    
\usepackage{hyperref}       
\usepackage{url}            
\usepackage{booktabs}
\usepackage{dsfont}
\usepackage{nicefrac}       
\usepackage{microtype}      
\usepackage{xcolor}         
\usepackage{graphicx}
\usepackage{subfigure}
\usepackage{booktabs} 
\usepackage{amsmath, amsfonts, bbm, bm} 
\usepackage[mathscr]{euscript} 
\usepackage[noend]{algorithm2e}
 \usepackage[
    backend=biber,
    style=numeric,
  ]{biblatex}

\newcommand{\bw}{\mathbf w}

\title{Cross Section Doppler Broadening prediction \\ using Physically Informed Deep Neural Networks}

\author{
  Arthur Pignet\\
  ENS Paris-Saclay, Gif-sur-Yvette, France \\
  Mines Paris, Paris, France \\
  \texttt{arthur.pignet@minesparis.psl.eu} \\
  \And
   Luiz Leal\\
  Institut de Radioprotection et de S\^{u}ret\'e Nucl\'eaire\\
  \texttt{luiz.leal@irsn.fr}
 \And
 Vaibhav Jaiswal\\
 Institut de Radioprotection et de S\^{u}ret\'e Nucl\'eaire\\
 \texttt{vaibhav.jaiswal@irsn.fr}
}

\begin{document}

\maketitle
\begin{abstract}
Temperature dependence of the neutron-nucleus interaction is known as the Doppler broadening of the cross-sections. This is a well-known effect due to the thermal motion of the target nuclei that occurs in the neutron-nucleus interaction. The fast computation of such effects is crucial for any nuclear application. 
Mechanisms have been developed that allow determining the Doppler effects in the cross-section, most of them based on the numerical resolution of the equation known as Solbrig's kernel, which is a cross-section Doppler broadening formalism derived from a free gas atoms distribution hypothesis. This paper explores a novel non-linear approach based on deep learning techniques. Deep neural networks are trained on synthetic and experimental data, serving as an alternative to the cross-section Doppler Broadening (DB). This paper explores the possibility of using \textit{physically informed neural networks}, where the network is physically regularized to be the solution of a partial derivative equation, inferred from Solbrig's kernel. The learning process is demonstrated by using the fission, capture, and scattering cross sections for $^{235}U$ in the energy range from thermal to 2250 eV.

\end{abstract}


\section{Introduction}
Cross-sections are closely linked to the probability of a given interaction to occur between particles. Therefore, their accurate and fast computations are essential for modern neutronics' simulations that involve multiphysics calculations. At non-zero temperature, the thermal agitation of the target nuclei influences the relative speed of the incident particle and the target, which has a strong impact on the neutron-nucleus interaction. This effect is known as Doppler Broadening effect, and can be modeled by a closed-form equation known as Solbrig’s kernel, given the assumption that the thermal distribution of nuclei follows a free-gas velocity distribution.

All cross-sections softwares store the cross-sections at 0K, whereas for practical applications higher temperature cross-sections cross sections are needed. Thus, the cross-sections at 0K need to be broadened. The calculation of the integral that appears in the Solbrig's kernel formalism, which can be expressed as a convolution over energies, is expensive, and its fast computation is a crucial issue. 

Various techniques have been introduced to compute the Solbrig’s Kernel in a fast and accurate way. 
For instance, in the BROADR module of NJOY \cite{osti_1338791}, the DB is carried out by integrating the Solbrig’s kernel. Another alternative consists of re-writing the Solbrig’s kernel as an equivalent heat-like equation on temperature and energy with the appropriate changes of variable \cite{Leal1987}. In the latter approach, it is proposed to solve a partial differential equation (PDE) by means of an explicit finite difference method. The energy-temperature meshes are selected in such a way that the second-order errors in the Taylor expansion in the mesh discretization vanish. The constraint in connection to the methodology is that energy-temperature meshes must be constant. This method is thus also intrinsically linear. In \cite{DBfourier2015}, the authors re-write the Solbrig’s kernel equation as a convolution product between a gaussian distribution and the un-broadened cross-sections. Using the fact that the Fourier basis diagonalizes the convolution operator \cite{mallatBook}, along with fast-fourier-transform algorithms, they introduced another linear method to compute the Solbrig's Kernel in a fast and accurate way by performing the convolution product in the frequency domain. 

The function mapping the incident particle's energy and the temperature to the corresponding cross-section being anything but regular, we believe that non-linear representations of this function should be sparser and computationally lighter. Motivated by the recent successes of deep neural networks (DNN) in various fields, which are by essence a class of non-linear approximators, we investigate the ability of such networks to represent broadened cross-sections. These learning methods are appropriate for highly non-linear representations while allowing an on-the-fly computation of the temperature-dependent cross sections at reasonable computational cost, as of that of a forward pass through a trained network. In addition, the memory cost of storing the network-trained weights is adequate. 

Even though the universal approximation property of deep neural network was proved a few decades ago \cite{HORNIK1989359}, it is only with recent advances in diverse mathematical fields, along with the increasing amount of available data and hardware power, that deep learning reaches ground-breaking performances in diverse domains, such as image analysis \cite{alexnet}, natural language processing \cite{bert}, or reinforcement learning \cite{dqn}. However, the massive amount of data needed to ensure proper results is prohibitive, especially in domains where data acquisition implies expensive experiments or simulations, as it is the case in fluid thermodynamic for instance. To overcome this deficiency, \textit{Physically Informed Neural Networks} (PINN) have been introduced, which consist of a class of neural networks trained in a supervised way, while being \textit{physically regularized}, what is to say constrained to approximate the solution of partial derivative equations \cite{raissi2019physics}.
In this paper, we investigate the ability of neural networks to reconstruct cross-sections with doppler broadening effect. We experiment two training processes on data generated with the NJOY modules RECONR and BROADR, for the fission, capture and scattering cross-sections of $^{235}U$ in the energy range from thermal to 2250eV. The first method consists in training a deep neural network in a supervised way to minimize the error to the data. The second one is to use PINN to regulate the network to approximate the solution of a PDE. The PDE is derived using the formalism proposed in \cite{Leal1987}. 
The method presented in this paper shows great promises as we obtain trained networks represented by 9,954 weights (32-bits floating point), able to reach 5.28 \% relative deviation in average. The computation of one pointwise cross-section takes only $1.7 10^{-5}$ seconds, no matter the temperature nor the energy, and can be heavily paralleled. 

\section{Doppler broadening effect}

The cross-sections of the neutron-nucleus interaction depend on the relative speed of the neutron and the target. Since the target is not at rest in the laboratory system due to thermal motion, a Doppler shift in the cross-section arises. The relation between the cross-section in the laboratory system and the effective cross-section can be expressed as follows:
\begin{equation}\label{eq:db}
 v \bar{\sigma}(\frac{mv^2}{2}) = \int p(\vec{W}) |\vec v - \vec W| \sigma(\frac{m|\vec v - \vec W|^2}{2})d^3W
\end{equation}
where $\bar{\sigma}$ is the Doppler broadened cross-section for incident particles of mass $m$ at speed $v$. The target's velocity distribution is denoted $p(\Vec{W})$. A common assumption is to model this distribution to be the same as in an ideal gas, i.e., the Maxwell-Boltzmann distribution. This hypothesis simplifies equation \ref{eq:db} into the following, known as the Solbrig's kernel equation: 
\begin{equation}\label{eq:solbrig}
 u^2\sigma_{\theta}(u) = \frac{1}{2\sqrt{\pi\theta}}\int_0^{\infty} u'^2\sigma_0(u')[e^{\frac{-(u-u')^2}{4\theta}} - e^{\frac{-(u+u')^2}{4\theta}}]du'
\end{equation}
where $u = \sqrt{E}$, $\theta = \frac{m k_b T}{2M}$, $m$ is the mass of the incidence particle and $E$ the energy. The temperature is denoted $T$, and $k_b$ is the Boltzmann's constant. 
However appealing this closed-form equation may be, it involves an integration all over the energies, which is computationally expensive. Following the work of \cite{DBfourier2015}, one can re-write this equation to exhibit a convolution product. First of all, we denote $\{r_\theta\}_{\theta\geq0}$ a family of functions defined by:
\begin{equation}
 r_\theta: u \rightarrow{} u|u| \sigma_\theta(u) 
\end{equation}
Using the parity of the involved functions, equation \ref{eq:solbrig} becomes:
\begin{equation}
 \forall u \in \mathbb{R},\quad r_\theta(u) = \frac{1}{2\sqrt{\pi\theta}} \int_{- \infty}^{+ \infty} r_0(u') e^{\frac{-(u'-u)^2}{4\theta}}du' = (r_0 * \frac{e^{\frac{-(.)^2}{4\theta}}}{2\sqrt{\pi\theta}})(u)
\end{equation}
where $*$ denote the convolution product.
One can notice that the operator $L_\theta: r_0 \rightarrow r_\theta $ is linear, and time-invariant. Therefore it is diagonal in the Fourier basis, which is thus the Karhunen-Lo\`eve basis, the linear optimal basis in which to approximate the operator $L$ \cite{mallatBook}. The cross-section being a signal with numerous very sharp and local fluctuations, we expect non-linear representations to be better than linear ones. 
We denote $F: \theta \rightarrow r_\theta$ the operator which maps the temperature of the target to the doppler-broadened cross-section. We overload the notation to denote $F(\theta)(u) = F(\theta,u)= r_\theta(u)$.
\subsection{Partial derivative Equation}
Deriving the Solbrig's kernel equation, it can be shown that F is the solution of a partial derivative equation \cite{Leal1987}; 
\begin{equation}\label{eq:PDE}
 \frac{\partial^2F}{\partial u^2} = 2\frac{\partial F}{\partial \theta}
\end{equation}
The method suggested in \cite{Leal1987} is to resolve this equation via finite element method, with a mesh of fixed-size. The mesh size is cleverly chosen to minimize the second-order term of the error induced by the finite-element scheme. In this paper, the PDE will be used to train a physically informed network. Actually, we will softly constrain the network to be the solution of this equation. 

\section{Artificial neural networks}

\subsection{Principle}
Neural networks are used in \textit{supervised learning}, where the objective is to approximate a function $f$, given a dataset $\mathbf D = {(x_i, y_i)}_{i<N}$ with $N$ samples, where $y_i = f(x_i)$. 
A feed-forward neural network $f_w$ (also called multi-layer perception) of parameter $\bw $ is said to be \textit{feed-forward} because the information flows from a given input vector $\mathbf{x} \in \mathbb{R}^d$ of dimension $d$, through the intermediate computations used to define $f_w$ , and finally to the output $f_w(x)$, without feedback loop. The intermediate computations are a succession of $K \in \mathbb{N}$ functions, called layers. Each layer is composed by a multivariate linear mapping, followed by a scalar wise non-polynomial function, called the activation function. We denote $\sigma$ the activation function, and $C_l, 0 < l < K$ the linear mappings. 
\begin{equation}
 \mathbf y= C_K \circ \sigma \circ C_{K-1} \circ \sigma \quad ... \quad \sigma \circ C_0 (x)
\end{equation}
Where 
\begin{equation} 
\forall l < K, C_l(\textbf{x}) = W_l\textbf{x} + b_l
\end{equation}
$W_l$ and $b_l$ are respectively called the weights and bias of the layer $l$. With a slight change of notation, we denote $\bw = \{W_l, b_l\}_{l<K}$ the parameters of the whole network. 
Many activation functions have been proposed in the last few years \cite{ramachandran2017searching}. Historically the sigmoid function had been used; nevertheless, it has widely been replaced by the \textit{rectified linear unit} (ReLu), which is the default choice nowadays. $C_K$ is called the output layer, $C_0$ is the input layer, and the other $C_l$ are referred to as the hidden layers. Networks with at least one hidden layer are called \textit{Deep} neural networks (DNNs). A major result is the universal approximation property of DNN \cite{HORNIK1989359}. Moreover, the intrinsic non-linearity of DNNs results in an ability to learn sparser representations of functions than polynomial approximations. However, the tremendous approximation power of DNN is not fully understood yet and is an active and open research topic.
The training or learning of a DNN is the optimization of its parameters $\bw$ to minimize a chosen error metric $\epsilon(f, f_w)$ between $f$ the function we aim to approximate and $f_w$ the representation expressed by the DNN. The training is usually done via stochastic gradient descent \cite{Goodfellow-et-al-2016} on an empirical approximation error, computed on a given dataset:
\begin{equation}
 \bw = \min_{w'} \sum_{i<N} \epsilon(y_i,f_{w'}(x_i))
\end{equation}
The optimized function is called the \textit{loss}. The gradient of the loss is efficiently computed via automatic differentiation, also referred to as gradient back-propagation algorithm \cite{Goodfellow-et-al-2016}. The reasons for the rise of deep learning strongly rely upon improvements in optimization techniques, loss choices, and the growth of usable datasets. 

\subsection{Physically Informed Neural Networks}

In applied physics, data is rare and expensive, as it often results from costly simulations or even the result of experiments that could not be well completed. To overcome these data issues, the authors of \cite{raissi2019physics} introduced the so-called \textit{Physically informed neural networks} (PINNs), with the main idea of benefiting from the vast amount of information encoded in known physical laws, expressed as partial derivative equations. A PINN is trained jointly by minimizing both the approximation error and the residual of the PDE. Considering parametrized and non-linear partial differential equations of the general form:
\begin{equation}\label{eq:genPDE}
 \frac{\partial f}{\partial t} + \mathcal{N}[f] = 0, u \in \Omega, t \in [0,T_max],
\end{equation}
where $f(u,t)$ denotes the solution of the PDE, $\mathcal{N}[.]$ is a non-linear differential partial operator and $\Omega$ is a subset of $f$'s definition set. We define the residual $R(u,t)$ to be the left-hand side of the previous PDE:
\begin{equation}
 R_f(u,t) = \frac{\partial f}{\partial t} + \mathcal{N}[f]
\end{equation}
Thus $f$ is solution of \ref{eq:genPDE} if and only if $R_f(u,t)=0, \quad \forall u \in \Omega, \forall t \in [0,T]$. We aim at approximating $f$ using a DNN $f_w$. The core idea in the work of \cite{raissi2019physics} is to jointly minimize the approximation error on points where $f(u,t)$ is known, i.e., the dataset $D$ (boundary conditions, but also experimental measures, and so on), and minimize $R_{f_w}(u,t)$, on another dataset $\tilde D = \{x_j\}_{0<j<\tilde N}$. The second minimization can be done on arbitrary points, as we do not need to know the exact value of $f$. Indeed our interest relies on the residual, therefore in the values of DNN's derivatives, which are computed via automatic differentiation \cite{Rumelhart1986}. The gradient of the residual with respect to the network's parameters $w$ is also computed as:
\begin{equation}
loss = \frac{1}{2N} \sum_{x,y \in D} \left\| f_w(x) - y \right\|_2 + \frac{1}{2\tilde N }\sum_{\tilde x \in \tilde D} \left\| R(x) \right\|_2 
\end{equation}
Here the chosen approximation error between $f$ and $f_w$ is the mean squared error, and $x = (u,t)$.

\section{Approximation of  Solbrig'S kernel}
\subsection{Datasets}
The objective of the paper is to approximate $F(u,\theta) = r_\theta(u)$ using a neural network. Our studies were conducted for the ${}^{235}U$ isotope in the resolved resonance region. The fission, capture, and scattering cross-sections in the energy range from thermal to 2250 eV were considered. Pointwise cross-sections were reconstructed with the NJOY modules RECONR and BROADR with a precision of $1 \%$ in steps of temperatures of 50 K starting from the unbroadened cross-section till 1000 K. The resulting training dataset is composed of 1,267,686 data points, for each kind of cross-sections. To evaluate the performance of our trained networks, we used cross-sections generated at the temperatures 335 K, 410 K, 505 K, 604 K, and 710 K, with the same sampling step along energies. It is worth noting that the cross-section data at these temperatures was never used in the network training and is nevertheless used to evaluate the performance and the ability of the trained networks to predict the temperature-dependent cross-sections generated by NJOY.

\subsection{Preprocessing}

Our experiments have shown that the approximation of the log of the cross-section works better than using the cross-section directly. Presently we do not have a clear picture of why the log of the cross-section is a better choice. More inspections are needed, which are left for future work. In the next section we will work with the function $\phi(u, \theta) = ln(F(u, \theta))$. Moreover, in order to have a balanced impact of the input, we normalize them and denote $\hat u$, and $\hat \theta$ the normalized input. 

In order to train the PINNs, we use a different preprocessing. We need a PDE for the function $\phi$. Starting from \ref{eq:PDE}, one can derive the following differential equation:
\begin{equation}
    \begin{split}
\frac{\partial^2 \phi}{\partial u^2} & = 2 \frac{\partial \phi}{\partial \theta} - (\frac{\partial \phi}{\partial u})^2 
\end{split}
\end{equation}
Eventually, we introduced a constant $\alpha$ to re-scale the derivatives, according to the scaling of the inputs. The exact scaling used will be discussed later. Approximating $\phi(\hat u,\hat \theta)$ with a deep DNN, $f_w$, we define the residual as:
\begin{equation}
R_w := \frac{\partial^2 f_w}{\partial \hat u ^2} -\alpha \frac{\partial f_w}{\partial \hat \theta} + (\frac{\partial f_w}{\partial \hat u})^2
\end{equation}
We train the PINNs' parameters by minimizing the loss:
\begin{equation}\label{eq:PINNloss}
 Loss =\frac{1}{N}\sum_{i=1}^{N} |f_w(\hat u _i,\hat \theta _i) - \phi^i|^2 + \frac{1}{N}\sum_{i=1}^{N_f}|R_w(\hat u _i,\hat \theta _i)|^2
\end{equation}

The minimization of the residual induces the existence of a trivial local minimum, where the network's derivatives are trivial everywhere, and the network itself yields the mean of the dataset output. To avoid this pitfall, we add a shortcut connection between the input $u$ and the output, i.e. $\phi(u, \theta)$  is approximated by $ f_w(u,\theta) + 2ln(u))$.
Using this shortcut connection imposes strictly positive inputs, which prevents the use of a standard scaling of our inputs. Moreover, the scaling of the energy contracts the signal, and makes the frequencies of the output signal (i.e. the cross section ) extremely high. Therefore the derivatives involved in the computation of the residual are also important, and make the residual blow up at the beginning of the training, which stops any attempt to train the network. On the other hand, as $\theta$ is valued around $10^{-5}$, it is out of the question to not scale at all the inputs, as one will be negligible compared to the other. Thus, when training PINN, we scaled only $\theta$, to be in the same range as $u$. However, in the case of the DNN, we scaled the inputs and the outputs to match a zero mean and standard variance.

\subsection{Network Architecture}
We use the same architecture of 10 layers and 32 neurons per layer for each cross section types, for both configurations, regular DNNs and PINNs. To ensure a proper back-propagation of the gradient, we used the weight normalization technique, introduced in \cite{salimans2016weight}. The design of networks' architecture is constrained by the use of PINN. The activation function chosen in all our experiment is not the widely spread \textit{Relu} \cite{relu}, which exhibits trivial first and second order derivatives, but the \textit{swish} activation function \cite{ramachandran2017searching}, which is defined as $ \forall x \in \mathbb{R}, \quad swish(x) = \frac{x}{1 + e^{-x}} $.  Note that with this specific architecture, one prediction costs 10 32 by 32 matrix products, one dot product on 32-elements vectors, and one 2 by 32 matrix-vector product. We also have 10 calls to the vectorized function $switch$. Moreover the network is deterministic, as there is no stochasticity involved in the \textit{prediction step}. Thus, multiple prediction with the same inputs will yields the same output. 

\textit{The training steps}, ie the optimization of the weights, is done through the popular gradient-based stochastic-first-order optimization algorithm or simply Adam \cite{kingma2017adam}. The initial learning rate is 1e-2, and it decreases when the loss reach a plateau. We noticed that, albeit this algorithm provides us with a fast learning, the resonance peaks of the function we aim to approximate were lost. We speculate that this is a consequence of the random noise added by the stochastic estimation of the gradient, and increased the batch size (from $2048$ to $5x10^6$) gradually throughout the optimization. This process helped us decrease the final error. 

Every hyper-parameter has been hand-selected through trials and errors. The use of hyper-parameters' systemic search algorithms is left for future work. To avoid inducing bias in our performance measurement by selecting the appropriate hyper-parameters for our given test set, we used 5\% of the training set, uniformly sampled, as a validation dataset. Not only is this dataset used to select the hyper-parameters, but also to avoid over-fitting, via early stopping, i.e., we stop the optimization when the error on the validation dataset increases, whereas the error on the training dataset is still decreasing.

\section{Numerical experiments}
\begin{table}
            \centering
            \begin{tabular}{|c||c|c|c|}
            \hline
               \textbf{Cross section} & \textbf{elastic} & \textbf{capture} & \textbf{fission} \\
                \hline
                        mean    &     2.49 & 5.52  &  5.78 \\
                min       &   1.7e-5 & 4.73e-5 &  7.9e-5 \\ 
                25 \%        &  0.76 & 2.1 &  2.0 \\
                50 \%         & 1.6 & 4.6 & 4.1 \\ 
                75 \%          & 3.2 & 7.7 & 7.5 \\ 
        \hline
    \end{tabular}
    \caption{Statistics errors (\%) for learning the cross sections within the energy range 200 eV - 300 eV using a DNN per cross section types.}
    \label{tab:error_stats}
\end{table}
 \begin{figure}[!ht]
    \centering
    \includegraphics[scale=0.4]{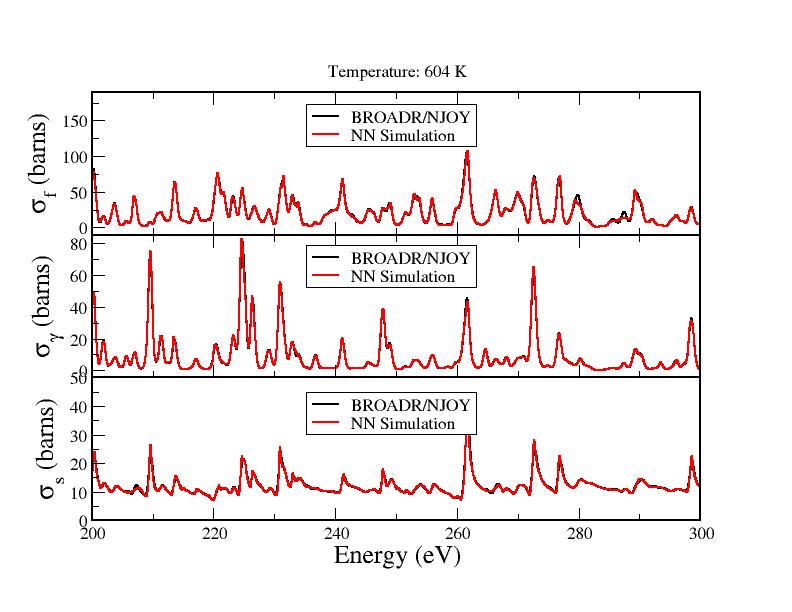}
    \caption{Fission, Capture, and Scattering cross sections calculated with BROADR and with a trained DNN.}
    \label{fig:xsec}
\end{figure}
This section reports on the performances achieved using the protocol mentioned before. All experiments were run via Google Colaboratory, on Tesla T4, or K80 GPUs. The results presented in Table \ref{tab:error_stats} demonstrate that DNNs are able to accurately represent broadened cross sections, using little memory and with adequate computational cost. These results also highlight the complexity of accurately learning fission and capture cross sections, whereas elastic cross sections reconstructions are very accurate. It must be recalled that only 9,954 parameters (30 kB for the full model) are used to reconstruct in less that 10 seconds the whole dataset, which represents more than 100,000 point-wise cross sections per second, with only 5.28 \% average relative error on the whole energy range, and 2.49 \% for an energy range between 200 and 300 eV. 

            
By incorporating the physical constraint in the DNN, the PINN approach, an improvement in the results is expected. However, it was observed for the present study that the results based on the PINN leave much to be desired in regard to the DNN for the entire energy range, that is, from 0 to 2000 eV. As a matter of fact, the PINN reconstruction error is twice that of the DNN. For example, for the elastic section, the DNN has an error of 5.28\%, while the PINN error cannot be reduced below 10.67\%. We are convinced that the reason for this effect is undoubtedly in the choice of the PDE, that is, the network was not able to minimize both the PDE residual and the mean squared error of the data. Indeed, we neglected a term in the convolution shown in Eq. \ref{eq:solbrig}, which impacts the PDE representation \cite{Leal1987}. The ${}^{235}U$'s cross sections have very high fluctuations of important amplitude. Therefore the truncation of the convolution neglects those high frequencies, and is a factor of error. According to this analysis we expect the PINN, which aims at verifying the PDE, to show very little ability at representing the high frequencies of the signal. 

Comparisons of fission, capture, and scattering cross sections calculated with BROADR and that using the procedure presented in this paper is displayed in Figure \ref{fig:xsec}.
The statistical errors related to the calculations shown in Fig. \ref{fig:xsec} are displayed in Table \ref{tab:error_stats}. It is perceived that there are consistently issues on reproducing the cross section at the resonance peak. This is in connection to the poor capacity to learn high frequencies comes from the stochastic noise of the gradient estimates, which tends to vanish when we use increasing batch size. Work on the improvement of a more accurate representation is underway, using more appropriate network architectures, and ensembling methods.  

\section{Conclusions}

The aim of the study presented in this paper was to investigate the use of DNN to account for the effect of temperature on the cross section. Different approaches have been investigated for the use of DNN. The resonance structure of $^{235}U$ (with very close resonance, interference effects, etc) is a very challenging problem for the DNN use. However, the results exhibits the high capacity of DNN to accurately reconstruct doppler broadened cross sections, while enabling on-the-fly computation at very adequate memory cost, with an error of 5.28 \% on an energy range of 2000 eV, and an error of 2.49 \% on a range of 100 eV. The networks are trained on a pure data-driven way. We also explored the possibility to training DNNs in a hybrid way, via physically informed neural networks for which no benefit has been observed. The apparent reason being the hypothesis used in the derivation of the partial differential equation for Doppler broadening. It is suggested that the derivation of the PDE could be less constrained as we do not need to solve it, neither analytically nor numerically. Work on this issue is underway. Moreover, we explored only the basis of layers and architectures of deep learning in this paper, and we are currently improving the accuracy of our representation using other network designs. 


\printbibliography

\end{document}